\documentclass[prd,showpacs,showkeys,twocolumn,floatfix]{revtex4}
\usepackage{bm}
\usepackage{graphicx}
\begin{document}
\title{On-brane data for braneworld stars}
\author{Matt Visser}
\email{matt.visser@vuw.ac.nz}
\homepage{http://www.mcs.vuw.ac.nz/~visser}
\affiliation{School of Mathematical and Computing Sciences,
Victoria University of Wellington, New Zealand}
\author{David L. Wiltshire}
\email{dlw@phys.canterbury.ac.nz }
\homepage{http://www.phys.canterbury.ac.nz/~physdlw/}
\affiliation{Department of Physics \& Astronomy, University of Canterbury,
Private Bag 4800, Christchurch, New Zealand}
\date{30 December 2002; Revised 20 February 2003;
\LaTeX-ed \today}
\begin{abstract}
  
  Stellar structure in braneworlds is markedly different from that in
  ordinary general relativity. As an indispensable first step towards
  a more general analysis, we completely solve the ``on brane''
  4-dimensional Gauss and Codazzi equations for an arbitrary static
  spherically symmetric star in a Randall--Sundrum type II braneworld.
  We then indicate how this on-brane boundary data should be
  propagated into the bulk in order to determine the full
  5-dimensional spacetime geometry. Finally, we demonstrate how this
  procedure can be generalized to solid objects such as planets.

\end{abstract}


\maketitle

\def\d{{\mathrm{d}}}
\def\be{\begin{equation}}
\def\ee{\end{equation}}
\def\E{{\mathcal E}}
\def\implies{\Rightarrow}
\def\frac#1#2{{\textstyle{#1\over#2}}}
\font\ser=cmssi10 \def\Zop{\hbox{{\ser Z\hskip-.4em Z}}}
\section{Introduction}

One physically important problem in the braneworld scenario is the
development of a full understanding of stellar structure and black
holes~\cite{wiseman,deruelle,maartens,casadio1,casadio2}. What is
already known is that stellar structure in braneworlds is rather
different from that in ordinary general relativity. The key point is
that if one is confined to making physical measurements on the brane,
then the restricted ``on-brane'' version of the Einstein equations
does not form a complete system for specifying the brane
geometry~\cite{shiromizu,garriga,RGK}. The problem lies in the fact
that the bulk 5-dimensional Weyl tensor feeds into the ``on-brane''
equations and so connects the brane to the bulk.

On the other hand, one may solve the ``on-brane'' version of the
Einstein equations to obtain a consistent class of boundary data that
satisfies the 4-dimensional Gauss and Codazzi equations~\cite{shiromizu},
analogously to the initial data in the standard $(3+1)$ decomposition
of globally hyperbolic manifolds in general relativity. Although the
relevant partial differential equations are now elliptic rather than
hyperbolic, one can use the boundary data in a completely analogous
fashion --- as input into the 5-dimensional bulk Einstein equations in
order to ``propagate'' the 4-dimensional geometry off the brane and
into the bulk \cite{wiseman,casadio2}.

In this Letter we present an algorithm for completely solving the
4-dimensional Gauss and Codazzi equations for a static spherically
symmetric star on the brane. This provides the most general boundary
data, suitable for then determining the bulk geometry by an
appropriate relaxation method~\cite{wiseman}.

Specifically, we will use Gaussian normal coordinates adapted to a
timelike 4-dimensional hypersurface (spacelike normal) in a
5-dimensional geometry,
\begin{equation}
{\d s_5}^2=\d\eta^2+g_{ab}\;\d x^a\;\d x^b
\label{GNC}
\end{equation}
and assume a type II Randall--Sundrum braneworld~\cite{RS2}, in which
the bulk metric is a 5-dimensional Einstein space, (\emph{i.e.}, with
Ricci tensor proportional to the metric), but the 4-dimensional
Lorentzian signature metric $g_{ab}$ is as yet undetermined.

If we impose $\Zop_2$ symmetry on this spacetime, and tune the
5-dimensional bulk cosmological constant and brane tension
appropriately, then the junction conditions together with the
projected 5-dimensional Einstein equations (the Gauss equations)
reduce to the induced but incomplete on-brane ``Einstein equations''
in the form~\cite{shiromizu,deruelle}
\begin{eqnarray}
G_{ab} &=& 8\pi\; T_{ab} - \Lambda_4 g_{ab} - \E_{ab}
\label{EinsteinOnBrane}\\
&&
- {\kappa^2\over4}
\left[
(T^2)_{ab} - \frac13 T \; T_{ab}
- \frac12 g_{ab} \left( T\cdot T - \frac13 T^2 \right)
\right].
\nonumber
\end{eqnarray}
where $\kappa^2$ is a constant inversely proportional to the brane
tension, $\Lambda_4$ is an on-brane cosmological term, the nonlocal
term $\E_{ab}$ is simply the projection of the 5-dimensional Weyl
tensor onto the brane, and in addition to the usual stress-energy term
there is also a nonlinear term quadratic in stress-energy. In addition
to (\ref{EinsteinOnBrane}) one finds that the extrinsic curvature
\begin{equation}
K_{ab} = {1\over2} {\partial g_{ab}\over\partial \eta},
\end{equation}
is related to the on-brane fields via
\begin{equation}
K_{ab} = - {\kappa\over2} \; \left[ T_{ab} -\frac13 g_{ab}T \right]
- {8\pi\over \kappa}\; g_{ab}.
\label{E:extrinsic}
\end{equation}
%
The Codazzi equation
\begin{equation}
\nabla_b [K^{ab} - K g^{ab}] = 0,
\label{Codazzi}
\end{equation}
is then equivalent to 4-dimensional stress-energy conservation for the
on-brane matter: $\nabla^a T_{ab} = 0$. Eqs.\ (\ref{EinsteinOnBrane})
and (\ref{E:extrinsic}) complete the specification of the boundary
data by effectively supplying the ``on brane'' metric and its normal
derivative.

The only truly general thing we know about the nonlocal Weyl tensor
projection term $\E_{ab}$ is that it is traceless, ${\E^a}_a=0$.
Together with (\ref{EinsteinOnBrane}) this implies that
\begin{equation}
R = - 8\pi \;T + 4 \Lambda_4 - 
\frac14\kappa^2\left[T\cdot T - \frac13 T^2\right].
\label{Rconstraint}
\end{equation}
In the vacuum region outside a star this reduces to
\begin{equation}
R = 4 \Lambda_4,
\end{equation}
whereas inside the star it takes the form
\begin{equation}
R = \hbox{[nonlinear-source]}.
\end{equation}
If for instance we are dealing with a perfect fluid then
\begin{equation}
R = 8\pi(\rho-3p) + 4\Lambda_4
+ \frac14\kappa^2
\left[
(\rho^2-3p^2) + \frac13 (\rho-3p)^2
\right].
\end{equation}
If we are restricted to making physical measurements on the brane,
then (\ref{Rconstraint}) is the only really general thing we can say.
Eqs.\ (\ref{EinsteinOnBrane}), (\ref{Rconstraint}) are \emph{much}
weaker than\ the 4-dimensional Einstein equations and so the solution
space will be much more general. Of course, we must also use the
extrinsic curvature constraint (\ref{E:extrinsic}) to complete the
boundary data, and then ultimately probe the bulk geometry off the
brane, and this will indirectly provide further restrictions.

Our aim here is to solve Eqs.~(\ref{EinsteinOnBrane}) and
(\ref{E:extrinsic}) in full generality for arbitrary static spherically
symmetric solutions on the brane, to provide appropriate data to
``propagate'' into the bulk \cite{wiseman} (via Eqs.\ (\ref{evolve})
below), to thereby test the consistency of candidates for realistic
stellar models.

\section{Vacuum [$\Lambda_4=0$]}

In this section we will consider the case when $\Lambda_4=0$.
For a braneworld star, in the vacuum region outside the surface
we are interested in solving $R=0$. In the usual way we can locally
adopt on-brane coordinates such that
\begin{equation}
\d s^2 = - \exp[-2\phi(r)]\;\d t^2 + {dr^2\over B(r)} + r^2 \d\Omega^2.
\end{equation}
If we now impose the condition $R=0$ we have one differential
constraint connecting two unknowns --- therefore there will be a
nondenumerable infinity of solutions parameterized by some arbitrary
function of $r$. Various specific cases have already been discussed
in the literature, but no attempt has been made at extracting the
general solution.
\subsection{Special cases}
\noindent
Known special case solutions include:\\
--- ``Reissner-Nordstr\"om-like'' ~\cite{maartens}
\begin{equation}
B(r) = \exp[-2\phi(r)] = 1 -{2M\over r} + {Z\over r^2}.
\end{equation}
Note that the parameter $Z$ is
\emph{not} an electric charge, but should be thought of as a tidal
distortion parameter.
\\
--- ``Spatial Schwarzschild'' ~\cite{casadio1}
\begin{equation}
B(r) =1 -{2M(1+\epsilon)\over r};
\end{equation}
\begin{equation}
\exp[-2\phi(r)] =
\left( \epsilon + \sqrt{\displaystyle1 -{2M(1+\epsilon)\over r} }\over
1+\epsilon \right)^2.
\end{equation}
(This geometry is also discussed in a rather different context in
ref.~\cite{dadhich}.)
\\
--- ``Temporal Schwarzschild''~\cite{maartens,casadio1}
\begin{equation}
B(r) = \left(1-{2M\over r}\right)
\left(1+{a\over r(1-{3\over2}{M\over r})}\right);
\label{tS1}\end{equation}
\begin{equation}
\exp[-2\phi(r)] =1 -{2M\over r}.
\label{tS2}\end{equation}
However these are all very specific special cases; and are in no way
general.

\subsection{General vacuum solution}
Let us first write the metric in the form
\begin{equation}
\d s^2 = 
- \;
\exp\left[-2\int_r^\infty g(\bar r) \; \d \bar r \right] \; \d t^2
+ {\d r^2 \over B(r)}
+ r^2 \; \d\Omega^2.
\label{E:metric}
\end{equation}
The function $g(r)$ is interpreted as the locally-measured acceleration
due to gravity; it is positive for a inward acceleration.

Now calculate the Ricci scalar $R$
\begin{equation}
R =
{
(2r+r^2 g) B' +(2 r^2 g^2 + 2 r^2 g' + 4 r g +2) B - 2
\over
r^2
}.
\end{equation}
For the vacuum case we set this equal to zero, which yields
\begin{equation}
(2r+r^2 g) B' +(2 r^2 g^2 + 2 r^2 g' + 4 r g +2) B - 2 = 0.
\end{equation}
If we treat $g(r)$ as input and view this as a differential equation
for $B(r)$, it is a first-order linear ODE, and hence explicitly
solvable. The integrating factor is
\begin{equation}
F(r;r_0) =
\exp\left\{ \int_{r_0}^r
{
1+ 2 \bar r g(\bar r) + \bar r^2 g(\bar r)^2 + \bar r^2 g'(\bar r)
\over
\bar r(1+ \bar rg(\bar r)/2)
} \; \d \bar r
\right\},
\end{equation}
where $r_0$ is any convenient reference point, and the general solution is
\begin{equation}
B(r) = \left\{
\int_{r_0}^r {F(\bar r;r_0)\over \bar r \left(1+\bar r g(\bar r) /2\right) }
\; \d \bar r + B(r_0) \right\} F(r;r_0)^{-1} .
\label{solB1}\end{equation}
Whereas $r_0$ is an arbitrary gauge parameter, the constant $B(r_0)$ is
related to physical parameters such as the mass and post-Newtonian
corrections. In the case of the ``temporal Schwarzschild''
solution (\ref{tS1}), (\ref{tS2}), for example, $B(r_0)=2(a+r_0)-3M$.

Alternatively one may use $B(r)$ as input to generate a first-order
ODE quadratic in $g(r)$ --- a Riccati equation. However, this has no
comparable general solution. The algorithm above is a modification of
Lake's construction for generating spherically symmetric perfect fluid
spacetimes~\cite{lake}. We suspect that an isotropic coordinate
version, based on~\cite{rahman}, may also be viable.

This general solution can be somewhat simplified by using integration
by parts on the integrating factor to remove the derivatives of
$g(r)$. We find
\begin{eqnarray}
F(r;r_0) &=&
{
\left(1+r g(r)/2\right)^2
\over
\left(1+r_0+g(r_0)/2\right)^2
}
\label{solF}\\
&&
\times\,\exp\left\{
\int_{r_0}^r 
{1+\bar rg(\bar r)+\bar r^2g(\bar r)^2
\over 
\bar r(1+\bar rg(\bar r)/2)} \; \d \bar r
\right\},
\nonumber\\
&=&
{
r \left[1+\frac12 r g(r)\right]^2
\over
r_0 \left[1+\frac12 r_0 g(r_0)\right]^2
} \;
F_2(r;r_0),
\end{eqnarray}
where we have defined
\begin{equation}
F_2(r;r_0) \equiv
\exp\left\{
\int^r_{r_0} {g(\bar r) [1+2\bar rg(\bar r)]
\over 
2(1+{1\over2}\bar rg(\bar r))} \;\d \bar r
\right\}.
\label{solF2}\end{equation}
Eq.\ (\ref{solB1}) then becomes
\begin{eqnarray}
B(r) &=& \Bigg\{
\int_{r_0}^r
\left(1+\bar r g(\bar r)/2\right)
F_2(\bar r; r_0)
\; \d \bar r
\nonumber\\
&&
 + r_0 \; \left[1+\frac12 r_0 g(r_0)\right]^2 \; B(r_0)
\Bigg\}
\nonumber\\
&&
\times \left\{ r \; \left[1+\frac12 r g(r)\right]^2 \;  
F_2(r;r_0)\right\}^{-1}.
\label{solB2}
\end{eqnarray}
This is now the explicit general (static spherically symmetric)
solution of the equation $R=0$ using the function $g(r)$ and the
constant $B(r_0)$ as arbitrary input. The construction is completely
algorithmic.

There is a potential subtlety if the Schwarzschild coordinate $r$ does
not increase monotonically with respect to the outward radial proper
distance, $\ell_r$. For ordinary stars in general relativity the
monotonicity of $r$ with respect to $\ell_r$ is guaranteed by the null
energy condition. In braneworld stars there is no particular reason to
believe in the monotonicity of $r(\ell_r)$, but our construction will
still hold piecewise on monotonic intervals.

\section{Stellar interior}

We now want to solve $R = S(r)$, with $S(r)$ a specified source. We
can now consider arbitrary values of $\Lambda_4$ without additional
complications by including a possibly non-zero $\Lambda_4$ in $S(r)$.
Proceeding exactly as above we find
\begin{eqnarray}
B(r) &=&
\left\{
\int_{r_0}^r
{1+ \bar r^2S(\bar r)/2\over \bar r \left[1+\bar r g(\bar r)/2\right] }
F(\bar r; r_0) \;\d \bar r + B(r_0)
\right\}
\nonumber\\
&&
\times
F(r;r_0)^{-1},
\end{eqnarray}
with the same integrating factor $F(r;r_0)$ of Eq.\ (\ref{solF}).
Equivalently
\begin{eqnarray}
B(r) &=&
\Bigg\{
\int_{r_0}^r \left(1+\frac12 \bar r^2S(\bar r)\right) 
\left(1+\frac12 \bar r g(\bar r)\right)
F_2(\bar r)\; \d \bar r
\nonumber
\\
&&
+ r_0 \; \left[1+\frac12 r_0 g(r_0)\right]^2 \; B(r_0)
\Bigg\} \;
\nonumber\\
&&
\times\left\{ r \; \left[1+\frac12 r g(r)\right]^2 \; F(r;r_0) \right\}^{-1}.
\end{eqnarray}
Once the source $S(r)$ is specified, this is fully general. In
addition one must specify an arbitrary function $g(r)$, and a single
arbitrary constant $B(r_0)$, and so algorithmically determine the
metric on the brane.

Of course the source $S(r)$ is somewhat restricted in that it is an
algebraic function of the on-brane stress-energy tensor, which is
itself restricted by 4-dimensional energy-momentum conservation
(equivalent to the Codazzi equation~(\ref{Codazzi}) as mentioned
above). For a static perfect fluid with spherical symmetry the
stress-energy has the form
\begin{widetext}
\begin{equation}
T^{ab} = \left[
\begin{array}{cccc}
\rho(r) \; \exp\left[+2\int_r^\infty g(\bar r) \; \d \bar r \right] &
0 &0 &0
\\
0 & p(r) \; B(r) &0 &0
\\
0 & 0 & {p(r) /r^2} & 0
\\
0 & 0 & 0 & {p(r) /( r^2 \sin^2\theta)}
\end{array}
\right].
\end{equation}
The equation of energy--momentum conservation gives
\begin{equation}
{d p\over dr} = - g(r) \; [\rho + p], \label{TOV}
\end{equation}
which, being a linear first-order ODE, has the exact closed-form
solution
\begin{eqnarray}
&&p(r) = \exp\left[-\int_{r_0}^r g(\bar r) \; \d \bar r \right]
\times
\left\{
p(r_0)
-
\int_{r_0}^r g(\bar r) \; \rho(\bar r) \;
\exp\left[+\int_{r_0}^{\bar r} g(\tilde r) \; \d \tilde r \right]
\; \d \bar r
\right\}.
\end{eqnarray}
\end{widetext}
The lesson now is that to find all possible conserved tensors $T^{ab}$
one is free to specify the function $\rho(r)$ and the number $p(r_0)$
arbitrarily, and thereby calculate $p(r)$, which now yields the full
tensor $T^{ab}$. From $T^{ab}(r)$ we now calculate $S(r)$, thereby
fixing the intrinsic geometry on the brane. By eq.~(\ref{E:extrinsic})
this also automatically generates the most general possible candidate
for the extrinsic curvature $K^{ab}$ compatible with the assumed
symmetries. It must be emphasised that only \emph{some} of these
braneworld geometries are physically meaningful, because one now needs
to extrapolate them off the brane to see if the ``graviton'' is still
bound~\cite{Shinkai}.

\section{Extrapolating off the brane}

The full algorithm is:
\begin{itemize}

\item Step 1: Pick an arbitrary density distribution $\rho(r)$ of
 matter inside the star; an arbitrary function $g(r)$; and a single
 number $p(r_0)$. Calculate $p(r)$ and so evaluate the stress-energy
 tensor $T^{ab}$ and the source term
\begin{equation}
S(r) = - 8\pi T + 4 \Lambda_4 -
\frac14\kappa^2
\left[
T\cdot T - \frac13 T^2
\right]
\end{equation}

\item Step 2: Armed with this source term $S(r)$ and the previously
 chosen $g(r)$, pick one additional number $B(r_0)$ in order to
 calculate the function $B(r)$ and thereby generate a candidate
 metric $g_{ab}$ for the on-brane physics.

 By calculating the 4-dimensional Einstein tensor for this candidate
 metric, one can rearrange Eq.~(\ref{EinsteinOnBrane}) to calculate
 $\E_{ab}=\,^{(5)}C_{\eta a\eta b}$ and so the find the projection of
 the five-dimensional Weyl tensor on the brane.

\item Step 3: Using this on-brane metric, and the on-brane
 stress-energy calculated in Step 1, use Eq~(\ref{E:extrinsic}) to
 calculate the extrinsic curvature $K_{ab}$.
 
\item Step 4: Evolve the metric off the brane. Sufficiently near the
  brane one can certainly use normal coordinates and so the standard
  result
\begin{equation}
{}^{(5)} R_{\eta a\eta b} = {\partial K_{ab} \over \partial \eta} +
K_{am} \; K^m{}_b
\end{equation}
applies. Rearrange this to yield:
\begin{equation}
{\partial^2 g_{ab}\over\partial \eta^2} = -{1\over2}\;
{g^{a'b'}} \; \;
{\partial g_{aa'}\over\partial \eta} \;
{\partial g_{b'b}\over\partial \eta}
+ 2 \; {}^{(5)} R_{\eta a\eta b}.
\label{evolve}
\end{equation}
\null
\smallskip

\noindent
Here $^{(5)}R_{\eta a\eta b}$ is calculable in terms of the projection
of the five-dimensional Weyl tensor and the bulk cosmological
constant.
\end{itemize}
This now provides a well-determined set of equations for extending the
on-brane metric into the bulk. Whether or not the resulting braneworld
model is actually viable depends on whether or not the bulk geometry
is ``well behaved'' --- in particular, is the graviton bound or
unbound~\cite{Shinkai}.

We must make the caveat that the Gaussian normal coordinates system
(\ref{GNC}) is likely to break down at some stage as one moves away
from the brane. This is not really a fundamental objection but more of
a technical issue. While it is easiest to set up the ``on-brane''
boundary conditions using Gaussian normal coordinates, as a practical
matter when it comes to numerically solving for the bulk geometry one
should be prepared to dynamically adjust the coordinate system.

\section{Solid planets}
%

If we are dealing with a situation of spherical symmetry that does not
correspond to a perfect fluid, such as a solid planet, the radial and
transverse pressures would be different and we would have
\begin{eqnarray}
R &=& 8\pi(\rho-p_r-2p_t) + 4 \Lambda_4
\\
&&+
\frac14\kappa^2
\left[
(\rho^2-p_r^2-2p_t^2) + \frac13 (\rho-p_r-2p_t)^2
\right].
\nonumber
\end{eqnarray}
The other major change arises in the on-brane stress-energy tensor,
which becomes
\begin{widetext}
\begin{equation}
T^{ab} = \left[
\begin{array}{cccc}
\rho(r) \; \exp\left[+2\int_r^\infty g(\bar r) \; \d \bar r \right] &
0 &0 &0
\\
0 & p_r(r) \; B(r) &0 &0
\\
0 & 0 & {p_t(r) /r^2} & 0
\\
0 & 0 & 0 & {p_t(r) /( r^2 \sin^2\theta)}
\end{array}
\right].
\end{equation}
Then covariant conservation gives the slightly more complicated
equation
\begin{equation}
{d p_r\over dr} = {2\over r}(p_t - p_r) - g(r) \; [\rho+p_r].
\end{equation}
This is still a linear first-order ODE, and has the exact closed-form
solution
\begin{eqnarray}
&&p_r(r) = r^{-2} \exp\left[-\int_{r_0}^r g(\bar r) \; \d \bar r \right]
\times
\left\{
\int_{r_0}^r 
\exp\left[+\int_{r_0}^{\bar r} g(\tilde r) \; \d \tilde r \right]
\left(2 \, \bar r p_t(\bar r) - g(\bar r) \; \rho(\bar r) \; \bar r^2\right ) 
\; \d \bar r
+ r_0^2 \; p_r(r_0)
\right\}.
\end{eqnarray}
\end{widetext}
The lesson now is that to find all possible conserved tensors $T^{ab}$
one is free to specify the functions $\rho(r)$, $p_t(r)$, $g(r)$, and
the number $p_r(r_0)$ arbitrarily, and thereby calculate $p_r(r)$
which now yields the full stress-energy tensor $T^{ab}$. This is now
used to calculate $S(r)$. Once $B(r_0)$ is specified the on-brane
intrinsic geometry (encoded in $B(r)$) is calculable. \emph{Ipso
facto}, this procedure also generates the most general possible
candidate for the extrinsic curvature $K^{ab}$ compatible with the
assumed symmetries. Thus spherically symmetric solid objects such as
planets are not much more difficult to deal with than are fluid
objects such as stars.

\section{Conclusion}

What we have done in this article is to find the most general set of
``on-brane data'' suitable for characterizing an arbitrary static
spherically symmetric braneworld star. This on-brane data is
characterized by two arbitrary functions $g(r)$, and $\rho(r)$, and
two arbitrary constants $B(r_0)$, and $p(r_0)$. This is an enormous
dataset, much more general than the various special cases discussed
in~\cite{maartens,casadio1,casadio2}. After suitable notational
modifications, this boundary data can be used as input into Wiseman's
relaxation algorithm for determining the bulk geometry~\cite{wiseman}.
Although Wiseman's algorithm~\cite{wiseman} is general, his numerical
examples were restricted to special choices of boundary data.
Hopefully, by characterizing the most general set of boundary data, we
are providing the ingredients for a more systematic treatment of the
numerical problem in future work.

Ultimately, it will be the bulk geometry that determines whether or
not a specific set of on-brane data leads to a physically meaningful
star~\cite{Shinkai}. We believe the technique we have developed here
could be quite powerful because it is completely algorithmic. Ideally,
we would hope that broad statements about the class of possible
stellar structure functions $S(r)$ could made. For example, by a
suitable characterization of possible pathological singularities in
the bulk geometry, it might be possible to rule out particular classes
of structure functions.

We conclude that braneworld stars are potentially \emph{much} more
complicated than standard general relativistic stars, and emphasise
that the coupling to the bulk will play an important role in
restricting the possible functions $g(r)$ which enter the
Tolman-Oppenheimer-Volkoff-like equation (\ref{TOV}).

\medskip

\noindent{\emph{Acknowledgement:}} This work was supported by the Marsden
Fund of the Royal Society of New Zealand.

\smallskip

\noindent{\emph{Note added:}} After this work was completed a paper by
Bronnikov and Kim appeared \cite{BK} which also derives the pure
vacuum solution (\ref{solF})--(\ref{solB2}), but in the context of
braneworld wormholes. That analysis does not consider the presence of
matter, nor does it consider the extrinsic curvature.


\end{document}